\documentclass[a4paper,10pt]{article}
\usepackage{graphicx}
\usepackage{amsmath}

\def\dbar{{\mathchar'26\mkern-12mu {\rm d}}}

\title{Nonequilibrium thermodynamics at the microscale:\\ work relations and the second law}
\author{Eliran Boksenbojm \\
Instituut voor Theoretische Fysica,\\ K.~U.~Leuven, B-3001 Leuven, Belgium \\ \\
Bram Wynants \\
Instituut voor Theoretische Fysica,\\ K.~U.~Leuven, B-3001 Leuven, Belgium \\ \\
Christopher Jarzynski \\
Department of Chemistry and Biochemistry \\
and Institute for Physical Science and Technology \\
University of Maryland, College Park, MD 20742 USA}

\begin{document}

\maketitle

\begin{abstract}
For macroscopic systems, the second law of thermodynamics establishes an inequality between the amount of work
performed on a system in contact with a thermal reservoir, and the change in its free energy.
For microscopic systems, this result must be considered statistically, as fluctuations around average behavior become substantial.
In recent years it has become recognized that these fluctuations satisfy a number of strong and unexpected relations,
which remain valid even when the system is driven far from equilibrium.
We discuss these relations, and consider what they reveal about the second law of thermodynamics and the nature of irreversibility
at the microscale.
\end{abstract}

\section{Introduction}

Looking back at the nineteenth century, we should not be surprised that the formulation of thermodynamics coincided with the industrial revolution.
Thermodynamics provides a deep and indispensable understanding of the principles by which engines and refrigerators operate, and the fundamental limits they must obey.
The development of thermodynamics in turn gave rise to statistical mechanics, as Boltzmann, Maxwell, Gibbs and others sought the link between concepts such as heat and temperature, and the Newtonian motion of atoms and molecules.
From these developments there emerged our modern understanding of thermodynamics as an effective theory for macroscopic systems, arising from a statistical treatment of their microscopic constituents.
The success of this approach rests the law of large numbers, which dictates
that the greater the number of constituents of the system, the smaller the relative size of deviations (fluctuations) from the average
behaviour. For macroscopic systems, significant deviations from the typical behaviour are virtually impossible, hence
computed averages coincide with measured values of the observables.

We are now in an era where the microscopic world is increasingly accessible, as evidenced by 
dramatic progress in areas of research such as single-molecule manipulation and nanotechnology.
Advances in experimental techniques, theoretical modeling and numerical simulation have produced an increasingly detailed understanding of biomolecular machines such as myosin, kinesin, and the complex apparatus responsible for the replication, transcription and translation of the genetic code in living cells \cite{how01}.
Moreover, laboratories around the world are using the techniques of supramolecular chemistry to synthesize molecular complexes with moving parts \cite{kay07}.
For such microscopic systems thermal fluctuations are centrally important, and it is no longer sufficient to focus our attention exclusively on average behavior.
This, together with natural scientific curiosity, has led to increased interest in asking what the laws of thermodynamics ``look like'', when applied to microscopic systems \cite{bus05}.

In this text, we address this issue, focusing in particular on the behavior of systems driven away from an initial state of thermal equilibrium.
We will discuss several results that pertain to such systems, and that involve the relationship between {\it work} and {\it free energy}.
While these results are in principle valid quite generally, in practice they are relevant mostly to microscopic systems, for which fluctuations are substantial.

As this text is a pedagogical discussion, we will start with a brief review of the relevant macroscopic thermodynamic relations,
using the familiar example of a stretched rubber band.
We will then consider how these relations might apply to a microscopic analogue: a single RNA molecule stretched with optical tweezers.
Here the laws of thermodynamics, particularly the second law, must be interpreted statistically, in terms of averages over many measurements.
However, it is the fluctuations around these averages that will be the object of central interest in this text:
as we will describe in detail in Section \ref{sec4}, these fluctuations -- which we might be tempted to dismiss as mere ``noise'' -- in fact satisfy a number of strong and unexpected relations, which remain valid even far from thermal equilibrium.

This text was prepared in large part by two of the authors (E.B. and B.W.), based on lectures given by the third author (C.J.) at the International Summer School on Fundamental Problems in Statistical Physics XII (FPSPXII), held in Leuven, Belgium in September, 2009.
It is meant to provide an introduction to the topics it covers, in a style that reflects the informal and stimulating atmosphere of the summer school.

\section{Brief review of thermodynamic processes}

\subsection{An experiment with a rubber band}

\begin{figure}[ht]
\begin{center}
 \includegraphics[width = 7cm]{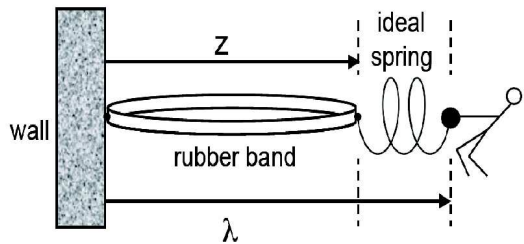}
\caption{Stretching a rubber band}
\label{rubber}
\end{center}
\end{figure}

Consider the system shown in Figure \ref{rubber}.
One end of a rubber band is attached to an immovable wall, the other end to a spring on which a force can be exerted. The length of the rubber band will be denoted by $z$ and the combined length of the rubber band and spring will be denoted by $\lambda$.
The system interacts with the surrounding air (thermal environment) which is at temperature $T$.
The parameter $\lambda$ will be controlled externally and will be used to exert work on the system. We assume that for every fixed value of 
$\lambda$ and $T$ there is a unique equilibrium state, to which the system relaxes if undisturbed.
 
We will use the term {\it thermodynamic process} to denote a sequence of events during which a system evolves from one equilibrium state to another.
During such a process, work ($W$) is performed on the system, 
heat ($Q$) flows into the system from the environment, and
the internal energy of the system may change ($\Delta U$).
The first law then reads
$$
\Delta U = W + Q
$$
There is some subtlety even in this seemingly straightforward statement of energy balance.
Namely, the definitions of $W$, $Q$ and $\Delta U$ depend on how we define our system of interest.  
For example, in Figure \ref{rubber} we could take the system of interest to be either the rubber band itself, or the rubber band together with the spring.
The corresponding definitions of work are then, respectively,
\begin{itemize}
\item system = rubber band alone : $W = \int F_{\textrm{spring}} dz$
\item system = rubber band + spring : $W = \int F_{\textrm{spring}} d\lambda$
\end{itemize}
where $F_{\textrm{spring}}$ is the instantaneous tension of the spring.
The difference between these two values is simply the net change in the energy of the spring itself.
Either choice is valid, but in this text we will use the second definition.

The second law of thermodynamics asserts that there exists a state function, {\it entropy} ($S$), that obeys the Clausius inequality,
$$
\int_{A}^{B} \frac{\dbar Q}{T} \leq \Delta S = S_B - S_A
$$
for any process that begins with the system in equilibrium state $A$ and ends in equilibrium state $B$.
Here the inexact differential $\,\dbar Q$ denotes an amount of heat absorbed from the thermal surroundings, at temperature $T$.
The equality is attained if and only if the process is reversible.

Let us now imagine the following process: initially the rubber band is at equilibrium
with the surroundings at temperature $T$ and the initial value $\lambda=A$ of the work parameter. Next we quickly stretch
the band by changing $\lambda$ to a new value $B$. During this process the rubber band will heat up a bit, so once the stretching is complete we let it cool down 
until it is again at equilibrium with the surroundings. For such a process, by combining the first and the second law we find :
\begin{eqnarray*}
 W & = & \Delta U - Q \\
 & = & \Delta U - T \int_{A}^{B} \frac{\dbar Q}{T} \\
 & \geq & \Delta U - T \Delta S
\end{eqnarray*}
Note that $T$ here denotes the (constant) temperature of the bath, rather than that of the system itself.
In terms of the Helmholtz free energy, $F = U - TS$, we have
\begin{equation}
\label{seclaw}
 W \geq \Delta F
\end{equation}
This inequality is the general statement of the second law for processes in which the system of interest is in contact with (at most) a single heat bath.
Since we will never consider the situation of two or more heat baths in this text, equation (\ref{seclaw}) is the form of the second law that we will use.

\subsection{Cyclic processes}

We will occasionally consider {\it cyclic} processes, that is, processes during which the value of the external parameter $\lambda$ is the same at the beginning as at the end.
For the rubber band example, one can imagine that after the stretching procedure described above, one brings the band back to its initial state, by varying $\lambda$ from
$B$ to $A$ and again allowing the system to equilibrate with its surroundings.

\begin{figure}[ht]
\begin{center}
 \includegraphics[width = 7cm]{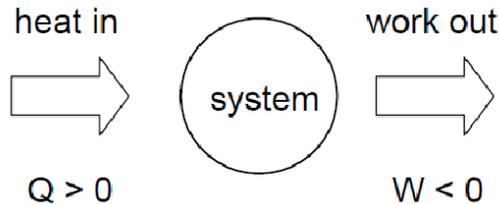}
\caption{The second law prohibits this situation, for cyclic processes.}
\label{free_lunch}
\end{center}
\end{figure}

For cyclic processes we have $\Delta F_{\textit{cyc}} = 0$ (since free energy is a state function)
thus equation (\ref{seclaw}) becomes
$$
W_{\textit{cyc}} \geq 0
$$
This is Thomson's formulation of the second law. \cite{tho1851}
Since the net change in the internal energy of the system is also zero during a cyclic process ($\Delta U_{\textit{cyc}}=0$) we must also have $Q_{\textit{cyc}}
 \leq 0$.
The opposite case ($W_{\textit{cyc}}<0<Q_{\textit{cyc}}$), illustrated schematically in Figure \ref{free_lunch}, is not possible for a cyclic process.
If it were, we could construct a so-called perpetual motion machine of the second kind, a device that would convert thermal fluctuations into work.
Thus Thomson's formulation is essentially a ``no free lunch'' statement: we cannot (for example) harvest the energy contained in the thermal motions of air molecules, and deliver that energy as work, simply by repeatedly stretching and contracting a rubber band.
(We stress that this conclusion is valid in the presence of a single thermal reservoir.
When two or more heat baths are present, we {\it can} obtain work simply by stretching and contracting of a rubber band, as illustrated beautifully in Ref. \cite{fey66}.)

Now let us split the cyclic process that takes $\lambda$ from $A$ to $B$ and back to $A$ into
a {\it forward} and a {\it reverse} process.
\begin{itemize}
 \item Forward process : $\lambda$ : A $\rightarrow$ B
 \item Reverse process : $\lambda$ : B $\rightarrow$ A
\end{itemize}
More precisely, let us assume the reverse process uses the time-reversed protocol for varying the work parameter: $\lambda_{t}^{R} = \lambda_{\tau-t}^{F}$,
where $\tau$ is the duration of either process.
If we now apply (\ref{seclaw}) to both the forward and the reverse process we find
\begin{eqnarray*}
 W_F \geq \Delta F & = &  F_B - F_A \\
 W_R  \geq - \Delta F & = & F_A - F_B ,
\end{eqnarray*}
equivalently
\begin{equation}
\label{eq:ineq_chain}
 -W_R \leq \Delta F \leq W_F
\end{equation}
Here too, the equalities are attained if and only if the process is reversible.
Equation (\ref{eq:ineq_chain}) immediately implies 
\begin{equation}
 W_F + W_R \geq 0
\end{equation}
which is simply Thomson's formulation of the second law.

\section{How might these results apply to microscopic systems?} \label{sec3}

\subsection{Example: stretching a strand of RNA}
\label{sec:RNA}

\begin{figure}[ht]
 \includegraphics[width = 10cm]{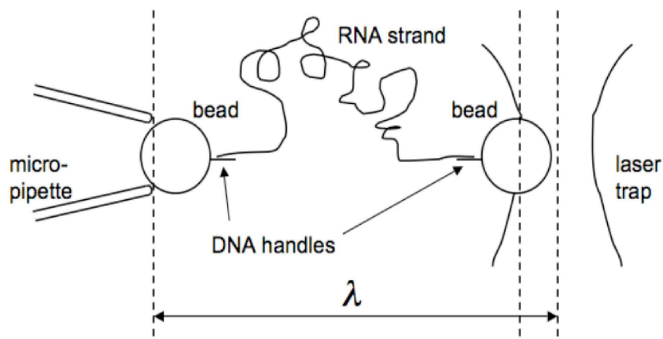}
\caption{Stretching a single molecule.
Note that this figure is not to scale: in reality the beads and DNA handles are much larger than the strand of RNA.}
\label{molecule}
\end{figure}

To illustrate how the thermodynamic results of the last section translate to
the microscopic world, it is again useful to consider a specific example.
Figure \ref{molecule} schematically depicts an experimental setup for stretching a single molecule, a microscopic analogue of the rubber band discussed in the previous section.
The system of interest here is a strand of RNA, which is chemically attached to a polystyrene bead, held fixed by a micropipette (analogous to the wall in
the rubber band example). The other end of the RNA is also attached to a bead, which is confined by
a laser trap (analogous to the spring in the previous example). The whole system is surrounded by 
water at room temperature, providing the thermal environment. By varying the position of the laser
trap ($\lambda$) work is performed on the system. 
For actual experimental implementations of such single-molecule manipulation experiments, within the context of the theoretical predictions of Section \ref{sec4}, see Liphardt et al. \cite{lip02}, Collin et al \cite{col05}, or (using atomic force microscopy instead of a laser trap) Harris and Song \cite{har07}.

\begin{figure}[ht]
 \includegraphics[width = 10cm]{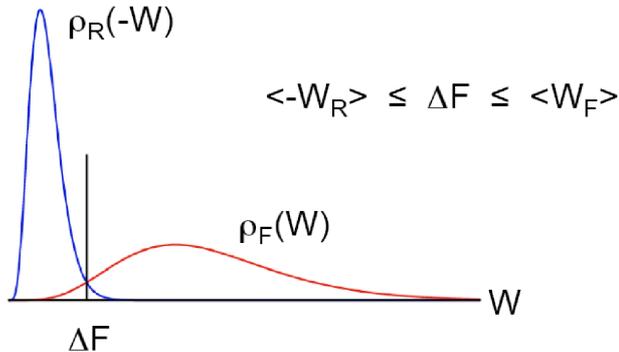}
\caption{Work distributions of the forward and reverse processes.}
\label{workdis}
\end{figure}

Now imagine the following irreversible process: start at $\lambda=A$, with the molecule in equilibrium with the thermal environment, then quickly stretch the molecule by varying $\lambda$ to a new value $B$, then let the system relax to a new equilibrium state, at $\lambda=B$.
Once this sequence of events is complete, change $\lambda$ back to 
$A$, following the reverse protocol, as in the previous section, and let the system relax again to equilibrium.
As above, let $W_F$ and $W_R$ denote the work performed during the forward (stretching) and reverse (contraction) stages, and $W_{\textit{cyc}} = W_F + W_R$.
If we repeat this cyclic process many times, we will not be surprised to measure different values of work, from one repetition to the next, as thermal fluctuations of the RNA strand and the surrounding water molecules play a significant role at this microscopic scale. We might even encounter rare occasions for which $W_{\textit{cyc}}<0$; these are in effect fortuitous events, during which the thermal motions
of the water molecules happen to facilitate the stretching and contraction of the RNA strand.
Because fluctuations are not negligible at this microscopic scale, we should interpret the second law as a statistical law, pertaining to expectation values, rather than a prediction about any particular  realization of an experiment.
Thus equation (\ref{eq:ineq_chain}) is replaced by the following statement about average work values:
\begin{equation}
\label{eq:avgineqchain}
\left<-W_R\right> \leq \Delta F \leq \left<W_F\right>
\end{equation}
Here and throughout this text, angular brackets denote an average over many realizations (repetitions) of the process.
For the full cycle we have
\[ \left<W_{cyc}\right> =  \left<W_F+W_R\right> \geq 0 \]
These considerations are illustrated in Figure \ref{workdis}, where 
work distributions $\rho_F(W)$ and $\rho_R(-W)$ for the forward and reverse processes are depicted.
We see that the average work values satisfy equation (\ref{eq:avgineqchain}), but each distribution shows a substantial spread of possible work values.
Note that the distributions have been drawn to intersect precisely at $W=\Delta F$; this is not a coincidence, but rather the consequence of equation (\ref{res2}) discussed in Section \ref{sec4} below.

\subsection{The second law from microscopic arguments}

The discussion in Section \ref{sec:RNA} can be summarized as follows: while the Clausius inequality might be violated during individual realizations of the process, $W < \Delta F$, it will be satisfied {\it on average}, $\left< W \right> \ge \Delta F$.
This conclusion is an intuitively natural way to extend the second law to microscopic systems, where fluctuations are non-negligible.
It is interesting to ask whether we can actually justify this intuitive conclusion using microscopic principles.
This is a deep question that has led to both insight and controversy, and we will not attempt to frame a satisfactory answer here.
Nevertheless, it is instructive to go through some simple analysis of an isolated Hamiltonian system, evolving under a cyclic process.
(In the discussion below, it is useful to keep in mind that this ``isolated system'' might be a proxy for the combination of a system of interest and a heat bath, if we consider the two together as a large, isolated system.)

Let the microstate 
$x = (q,p) = (\vec{q}_1,\ldots,\vec{q}_m,\vec{p}_1 \ldots,\vec{p}_m )$ describe the positions and
momenta of all the particles in the system. The $6m$-dimensional space that $x$ belongs to is the
phase space $\Gamma$. The Hamiltonian $H_{\lambda}(x)$ is a function of this microstate and one or more parameters $\lambda$,
and governs the dynamics by Hamilton equations:
\[ \dot{q} = \frac{\partial H}{\partial p}, \ \ \ \ \ \ \ \dot{p} = -\frac{\partial H}{\partial q} \]
This describes a fully deterministic dynamics, i.e. when $x(0)$ is given, $x(t)$ is known for any later
time $t$. This $x(t)$ (or $x_t$) defines what we call a trajectory or path through the phase space.

Hamiltonian dynamics satisfy Liouville's theorem: take a subset $C$ of the phase space, with volume $|C|$. 
If we let all the points $x\in C$ evolve under Hamiltonian dynamics to a later time $t$, this defines
a new subset $C_t = \{ x(t)| x(0)\in C \}$ with volume $|C_t|$. Liouville's theorem then states that $|C_t| = |C|$,
or equivalently:
\[ \left| \frac{\partial x_t}{\partial x_0}\right| \equiv \det\left( \frac{\partial x_i(t)}{\partial x_j(0)} \right) = 1  \]
For an isolated system, the first law of thermodynamics dictates that
\[ W = \Delta U = H_B(x_t) - H_A(x_0) \]
where again $A$ and $B$ denote the values of the external parameter $\lambda$ at the beginning and the end of the evolution.
For a cyclic evolution of the parameter, $A=B$.
We can now pose the question:
does the inequality $W_{\textit cyc} \ge 0$ hold for every initial condition $x(0)$?
In other words, for this isolated system is Thomson's law satisfied for every trajectory?
Without attempting a rigorous analysis, the following simple argument suggests the answer is ``no''.

\begin{figure}[ht]
 \includegraphics[width = 10cm]{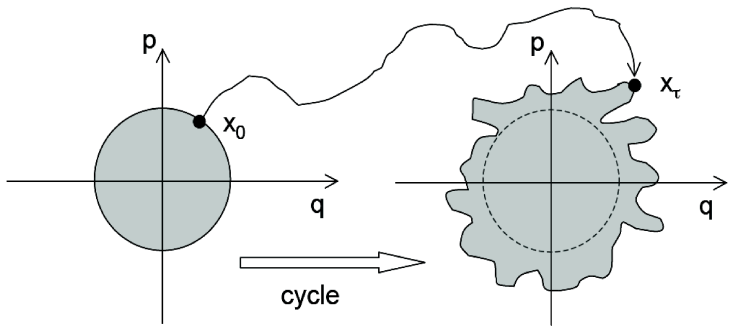}
\caption{Hypothetical phase diagram illustrating the scenario $C \subset C_\tau$.}
\label{liouville}
\end{figure}

For simplicity, imagine a Hamiltonian $H_A(x)$ with a single minimum in phase space, and consider the set of initial conditions $C$ corresponding to all microstates with an energy lower than
some value $U_0$, see Figure \ref{liouville} (on the left).
Let $C_\tau$ denote the set of final conditions reached after Hamiltonian evolution during a cyclic process.
There are three scenarios for the relationship between these two subsets of phase space: $C \not\subseteq C_\tau$, $C \subset C_\tau$, and $C = C_\tau$.
The first implies that there exists some point $\bar x$ that belongs to $C$ but not to $C_\tau$.
If we now consider a trajectory that ends at this point, then by construction the initial and final conditions for this trajectory satisfy:
\begin{equation}
x_0 \notin C \quad,\quad x_\tau = \bar x \in C
\end{equation}
Hence $W_{\textit cyc} < 0$ for this trajectory.
The second scenario ($C \subset C_\tau$) is depicted on the right side of figure \ref{liouville}, and implies $|C_\tau|\geq|C|$, a clear contradiction with Liouville's theorem; therefore we can rule out this scenario.
Finally, the third scenario, namely $C = C_\tau$, corresponds to the limiting case of a reversible process, for which we expect $W_{\textit cyc}=0$.
Thus we conclude that (unless the process is performed reversibly) there will exist initial conditions for which $W_{cyc} < 0$.

The above considerations suggest that the second law ought to be interpreted statistically: 
\[ \left<W_{cyc}\right> = \int dx_0 f_A(x_0)\left[ H_A(x_t(x_0)) - H_A(x_0) \right] \geq 0 \]
where $f_A(x_0)$ is a normalized probability distribution that represents the statistical state of equilibrium at $\lambda=A$.
Whether or not this inequality is satisfied depends, of course, on the choice of distribution $f_A$.
In the following paragraph we show below that the inequality is satisfied for a canonical distribution of initial conditions, $f_A(x) \propto e^{-\beta H_A(x)}$. 
It is also straightforward to establish that the inequality is satisfied if $f_A(x) \propto \theta(U_0-H_A(x))$,
where $U_0$ is a constant and $\theta(\cdot)$ is the unit step function.
(We leave this as an exercise for the reader.)
Somewhat surprisingly, however, for a microcanonical distribution of initial conditions,
$f_A(x) \propto \delta(U_0-H_A(x))$, it is possible to construct explicit examples for which $\langle W_{cyc}\rangle < 0$ \cite{mar09,sat02}.

Consider a map $M: \Gamma \to \Gamma: x \mapsto y$, for which $\left|{\partial y}/{\partial x}\right| = 1$ 
and for which the inverse $M^{-1}$ exists.
An example of such a map is Hamiltonian evolution for some fixed time ($x = x_0$, $y = x_\tau$).
Define furthermore a positive constant $\beta$ and the work $W(x) := H_A(y(x))-H_A(x)$. 
We compute the following average:
\begin{eqnarray}
 \left<e^{-\beta W}\right> &=& \int dx \frac{1}{Z_A}e^{-\beta H_A(x)}e^{-\beta W(x)}\nonumber\\
&=& \frac{1}{Z_A}\int dx \,e^{-\beta H_A(y(x))}\nonumber\\
&=& \frac{1}{Z_A}\int dy \left|\frac{\partial x}{\partial y}\right| e^{-\beta H_A(y)}\nonumber\\
&=& 1 \label{thomson}
\end{eqnarray}
where $Z_A$ is the usual partition function.
From Jensen's inequality (see Appendix A for a brief derivation) it follows that
\[ 1 = \left<e^{-\beta W}\right> \geq e^{-\beta\left<W\right>} \]
so that indeed $\left<W\right>\geq0$.
A more general version of this derivation can be found in e.g. \cite{jar02,obe05}.

These very simple considerations should not be viewed as first-principles derivations of the second law.
Rather, they provide some quantitative justification for the intuition stated at the beginning of this section: the second law (which in this case takes the form $W_{cyc}\ge 0$) will be violated for individual trajectories, but will be satisfied for an average over an ensemble of trajectories, at least for certain choices for the distribution of initial conditions.

\section{Nonequilibrium work and fluctuation relations}\label{sec4}

The central point of Section \ref{sec3} above is that -- for microscopic systems -- the second law ought to be interpreted statistically, as a statement about averages.
We now turn to the main focus of this contribution, namely the {\it fluctuations} around these averages.
We will discuss three prediction, given by equations (\ref{res1}), (\ref{res2}) and (\ref{res3}) below, that pertain to these fluctuations.
After introducing these results, we will derive each of them, using three different schemes for modeling the microscopic dynamics of a system driven away from equilibrium.
In each case, we imagine (as with the single-molecule stretching in Section \ref{sec:RNA} above) that the system of interest is prepared in equilibrium with a heat bath, then driven away from equilibrium by varying the work parameter over a time interval $0<t<\tau$, from $\lambda_0=A$ to $\lambda_\tau=B$.

\begin{enumerate} 
\item The first prediction is the {\it nonequilibrium work relation} \cite{jar97a},
\begin{equation}\label{res1}
 \left< e^{-\beta W} \right> = e^{-\beta \Delta F}
\end{equation}
As before, $W$ is the work performed during a single realization, $\Delta F = F_B - F_A$ is the free energy difference between the initial and final equilibrium states, $\beta = (k_BT)^{-1}$ is the inverse temperature of the heat bath, and angular brackets denote an average over an ensemble of realizations of the process.
In Section \ref{sec:hamiltonian} below, we will derive equation (\ref{res1}) under Hamiltonian dynamics, and we will show that this equality immediately implies two inequalities related to the second law:
\begin{itemize}
 \item $\left< W \right> \geq \Delta F$
 \item ${\rm Prob}(W \leq \Delta F - n\, k_BT) \leq e^{-n} $ 
\end{itemize}
for any real, dimensionless $n > 0$.
\item The second prediction, a {\it fluctuation theorem} due to Crooks \cite{cro98,cro99},
\begin{equation}\label{res2}
 \frac{\rho_F(W)}{\rho_R(-W)} = e^{\beta (W - \Delta F)}
\end{equation}
will be proven in Section \ref{sec:stochastic} for stochastic jump processes. 
Note that equation (\ref{res1}) follows directly from equation (\ref{res2}).
Moreover, this result implies that the
work distributions for the forward and reverse processes cross one another at $W=\Delta F$, as depicted in Figure \ref{workdis}.
\item 
Equations (\ref{res1}) and (\ref{res2}) in principle allow us to determine an equilibrium free energy difference from measurements of nonequilibrium work values.
The third prediction goes further, giving a prescription for constructing the entire equilibrium distribution \cite{hum01,hum05}: 
\begin{equation}\label{res3}
\left< \delta(x-x_t) e^{-\beta W_t} \right> = \frac{1}{Z_A} e^{-\beta H(x, \lambda_t)}
\end{equation}
where $W_t$ is the work done up to time t for a trajectory $x_t$ and $Z_A$ is the partition function at time $t=0$.
(Equivalent formulations of this result appear in Refs. \cite{jar97b,cro00}.)
We will derive this result for the case of continuous time Markov processes.
\end{enumerate}

Before proceeding to the derivations of these results, some general remarks are in order.
First, these results remain valid even if the system is driven far from equilibrium during the process.
They place strong and rather unexpected constraints on the work fluctuations: one would not have naturally guessed the validity of equations (\ref{res1}) - (\ref{res3}) simply by extrapolating from macroscopic, thermodynamic intuition.
Next, these results reveal that information about equilibrium properties (such as $\Delta F$) is in effect encoded in the fluctuations of the system when driven out of equilibrium.
Finally, equations (\ref{res1}) - (\ref{res3}) are just three results among a considerably larger set of predictions related to fluctuations in far-from-equilibrium processes.
In particular, the early work of Bochkov and Kuzovlev \cite{boc77,boc81a,boc81b} includes related predictions, using a somewhat different definition of work.
Furthermore, while we consider only fluctuations in {\it work}, there has been extensive investigation of fluctuations in {\it entropy production}, which satisfy similar relations; see e.g.\ Refs. \cite{eva93,eva94,gal95,kur98,leb99,mae99,mae03,har07b,sev08}.

\subsection{Hamiltonian dynamics}
\label{sec:hamiltonian}

Here we derive equation (\ref{res1}), using Hamiltonian dynamics to model the microscopic evolution of the system.
Specifically, we imagine that after the system has been prepared in equilibrium, at $\lambda=A$, the heat bath is removed and the now-isolated system evolves under Hamilton's equations as the work parameter is varied from $A$ to $B$.
At the end of the process the system is again brought into thermal contact with the heat bath and allowed to reach equilibrium.
This scenario does not realistically describe many physical situations, such as the RNA stretching experiment, therefore the derivation of equation (\ref{res1}) that we present here is not the most general one.
However, it is a particularly simple derivation, which avoids some of the technical complications that arise when treating a system that remains in contact with a heat bath during the entire process;
see Ref.~\cite{jar04} for a Hamiltonian analysis that explicitly includes the degrees of freedom of the system and the heat bath.

Our system is described by a Hamiltonian $H(x,\lambda)$, where $x = (\textbf{q}\: , \textbf{p})$.
Because the system is isolated, the work performed as the external parameter is varied from $\lambda_0=A$ to $\lambda_\tau=B$ is 
\begin{equation*}
 W = H(x_{\tau}, B) - H(x_0, A)
\end{equation*}
Since Hamiltonian dynamics are deterministic, we can view the final conditions as a function of the initial conditions, thus writing
\begin{equation}
W = H(x_{\tau}, B) - H(x_0, A) = H(x_{\tau}(x_0), B) - H(x_0, A) = W(x_0) 
\end{equation}
The only stochasticity is in the assignment of initial conditions for a given realization, therefore to compute the left side of equation (\ref{res1}) we must average over the distribution of these conditions:
\[ \left< e^{-\beta W} \right> = \int dx_0 f^{\textit{eq}}(x_0, A)e^{-\beta W(x_0)} \]
Here $f^{\textit{eq}}$ is the equilibrium distribution, given by the canonical expression
\begin{equation}
 f^{\textit{eq}}(x, \lambda) = \frac{1}{Z_{\lambda}} e^{-\beta H (x, \lambda)}
\end{equation}
We now perform the averaging as in the previous section:
\begin{eqnarray*}
 \left< e^{-\beta W} \right> & = & \int dx_0 \frac{1}{Z_A} e^{- \beta H(x_0, A)} e^{- \beta W(x_0)}\\
 & = & \frac{1}{Z_A} \int dx_0 e^{-\beta H(x_{\tau}(x_0), B)} \\
 & = & \frac{Z_B}{Z_A} = e^{- \beta \Delta F}
\end{eqnarray*}
where the free energy is given by the familiar expression,
\begin{equation}
 F(\lambda) = - \frac{1}{\beta} \ln Z_{\lambda}\ \ \ \ \ \ \ \ \Delta F = F(B) - F(A)
\end{equation}

One consequence of (\ref{res1}) is that the second law (in the form of the Clausius inequality) appears as an application of Jensen's inequality:
since the exponential function is convex, we have that $\left< e^{-\beta W} \right> \geq e^{-\beta \left<W\right>}$, 
and therefore $\left< W\right> \geq \Delta F$.
We can also use equation (\ref{res1}) to establish a bound on the probability to observe work values that ``violate'' the second law, $W < \Delta F$, as follows.
If $\rho(W)$ represents the probability distribution of work values, then the probability that the work is smaller than $\Delta F - nk_BT$, where $n>0$, is
\begin{eqnarray*}
{\rm Prob}(W\leq \Delta F - nk_BT) &=& \int_{-\infty}^{\Delta F - nk_BT} dW \rho(W)\\
&\leq&  \int_{-\infty}^{\Delta F - nk_BT} dW \rho(W)e^{-\beta(W-\Delta F + nk_BT)}\\
&\leq& e^{\beta(\Delta F - nk_BT)}\int_{-\infty}^{+\infty}dW\rho(W)e^{-\beta W}\\
&=& e^{-n}
\end{eqnarray*}
In other words the probability to observe work values smaller than $\Delta F$ decays exponentially, or faster, in the size of the violation.
This is consistent with our experience that macroscopic violations of the second law are never observed.

\subsection{Stochastic jump processes}
\label{sec:stochastic}

Here we derive equation (\ref{res2}) for a discrete-time Markov dynamics. 
Instead of representing the evolution of the system as a continuous trajectory, we imagine a chronologically ordered sequence of configurations,
\begin{eqnarray*}
 x_0 \rightarrow x_1 \rightarrow x_2 \rightarrow \cdots \rightarrow x_{\tau - 1} \rightarrow x_{\tau}
\end{eqnarray*}
which might represent ``snapshots'' of the system at equally spaced intervals of time.
We imagine that the system remains in contact with the heat bath throughout this process, therefore the transitions from one state to the next are stochastic.

These dynamics are specified by the transition probability densities $P_{\lambda} (x \rightarrow x')$, which depend on time through the parameter $\lambda$, and which satisfy the normalization condition
\begin{equation}
 \int dx' P_{\lambda} (x \rightarrow x') = 1
\end{equation}
We assume that the process is Markovian, in other words each transition is statistically independent of the previous transitions.
We also assume that each transition satisfies a detailed balance condition
\begin{equation}\label{detbal}
 \frac{P_{\lambda} (x \rightarrow x')}{P_{\lambda} (x \leftarrow x')} = e^{-\beta [H_{\lambda}(x') - H_{\lambda}(x)]}
\end{equation}
where the direction of the arrow indicates the transition.
We can interpret this condition by imagining for a moment that $\lambda$ is held fixed, allowing the system to relax to a state of equilibrium represented statistically by the Boltzmann distribution.
Equation (\ref{detbal}) then tells us that in this state the net rate of transitions from $x$ to $x'$ is the same as that from $x'$ to $x$:
\begin{equation}
 \frac{1}{Z_{\lambda}} e^{-\beta H_{\lambda}(x)} P_{\lambda} (x \rightarrow x') = \frac{1}{Z_{\lambda}} e^{-\beta H_{\lambda}(x')} P_{\lambda} (x \leftarrow x')
\end{equation}

Of course, we are interested in a process during which the value of $\lambda$ is not held fixed, and therefore the system does not (in general) remain in equilibrium.
To describe this situation, let us imagine that $\lambda$ changes in between each transition.
We represent this situation schematically as follows:
\begin{eqnarray*}
\lambda_0 = A & x_0 \xrightarrow{\lambda_1} x_1 \xrightarrow{\lambda_2} x_2 \xrightarrow{\lambda_3} \cdots \xrightarrow{\lambda_{\tau -1}} x_{\tau -1} 
\xrightarrow{\lambda_{\tau}} & \lambda_{\tau} = B 
\end{eqnarray*}
Following Crooks \cite{cro98}, now establish a link with thermophysics by specifying what we mean by heat and work in this model.
We define heat as
\begin{equation}\label{heat}
 Q = \sum\limits_{t=1}^{\tau} \left[ H(x_t, \lambda_t) - H(x_{t-1}, \lambda_t) \right]
\end{equation}
which is the sum of energy changes due to Markov transitions, while the work is defined as
\begin{equation}
 W = \sum\limits_{t=1}^{\tau} \left[ H(x_{t-1}, \lambda_t) - H(x_{t-1}, \lambda_{t-1}) \right]
\end{equation}
which is the sum of energy changes due to the changing of $\lambda$.
Taking the sum of these two contributions, we arrive at a statement of the first law of thermodynamics:
\begin{equation}
 Q + W = H(x_{\tau}, B) - H(x_0, A)
\end{equation}

Consider now a forward protocol and its reversed protocol,
\begin{eqnarray*}
 A = \lambda_0^F \rightarrow \lambda_1^F \rightarrow \cdots \rightarrow \lambda_{\tau}^F = B \\
A = \lambda_{\tau}^R \leftarrow \lambda_{\tau-1}^R \leftarrow \cdots \leftarrow \lambda_{0}^R = B
\end{eqnarray*}
related by $\lambda_{t}^{F} = \lambda_{\tau - t}^{R}$. Likewise we define a forward and backward trajectory,
$X = (x_0 \rightarrow x_1 \rightarrow \cdots \rightarrow x_{\tau})$
and
$X^{\dagger} = (x_0 \leftarrow x_1 \leftarrow \cdots \leftarrow x_{\tau})$. 
If $P^F(X)$ denotes the probability density for generating the trajectory $X$ during the forward process, and $P^R(X^\dagger)$ is defined analogously, then their ratio is
\begin{equation}
 \frac{P^{F}(X)}{P^R(X^{\dagger})} = \frac{f_{A}^{\textit{eq}}(x_0) P_{\lambda_1^F} (x_0 \rightarrow x_1) \cdots P_{\lambda^F_{\tau}} (x_{\tau -1} \rightarrow x_{\tau})}{f_{B}^{\textit{eq}}(x_{\tau}) P_{\lambda^R_0} (x_{\tau} \rightarrow x_{\tau-1}) \cdots P_{\lambda^R_{\tau-1}} (x_{1} \rightarrow x_{0})}
\end{equation}
with initial conditions for either process sampled from the corresponding Boltzmann distribution.
Combining the assumption of detailed balance (\ref{detbal}) with the definition of heat (\ref{heat}), we find that
\[ \frac{P^{F}(X)}{P^R(X^{\dagger})} = \frac{f_{A}^{\textit{eq}}(x_0)}{f_{B}^{\textit{eq}}(x_{\tau})}e^{-\beta Q^F}\]
where $Q^F$ is the heat dissipated during the forward path ($Q^F = -Q^R$). Using the explicit expressions for the canonical distributions,
we arrive at \cite{cro98}
\begin{equation}\label{tussenres} \frac{P^{F}(X)}{P^R(X^{\dagger})} = e^{-\beta \Delta F + \beta \Delta U - \beta Q^F} = e^{\beta W^F-\beta \Delta F} \end{equation}
where $W^F = W^F (X)$ is the work done during the forward process. 
With this expression, deriving equation (\ref{res2}) is simple:
\begin{eqnarray*}
 \rho_F (W) & = & \int d X P^{F} (X) \delta (W - W^F (X)) \\
 & = & \int d X P^R (X^{\dagger}) e^{\beta (W^F (X) - \Delta F)} \delta (W - W^F (X)) \\
 & = & e^{\beta (W - \Delta F)} \int d X^{\dagger} P^{R}(X^{\dagger}) \delta (W + W^R (X^{\dagger}))\\
 & = & e^{\beta (W - \Delta F)} \rho_R (-W)
\end{eqnarray*}
where $dX = dx_0 dx_1 \cdots dx_{\tau} = dX^{\dagger}$, and we have used $W^F(X) = -W^R(X^{\dagger})$.

From (\ref{res2}), we see that $\rho_F(\Delta F) = \rho_R(-\Delta F)$, meaning that observing a work value equal to the free energy differences is equally (un)likely during either process.
Furthermore, rewriting equation (\ref{res2}) as
\[ \rho_F (W)e^{-\beta W} = \rho_R(-W)e^{\beta \Delta F} \]
then integrating over $W$, we recover equation (\ref{res1}).

\subsection{Continuous time Markov dynamics}
\label{sec:feynmankac}

Equation (\ref{res3}) is essentially an application of the Feynmann-Kac theorem, as emphasized by Hummer and Szabo \cite{hum01,hum05}.
Let us therefore first sketch a (non-rigorous) derivation of this theorem in the context of continuous time Markov processes.
For a more mathematically rigorous approach, see Ge and Jiang~\cite{ge09}.

\paragraph*{Preliminaries:}
We define $P\left[X|x_0,t_0\right]$ to be the probability density
to observe the trajectory $X$, given that the system was in state $x_0$ at time $t_0$.
When we integrate this over all possible trajectories that end in state $x$ at time $t$,
we get the probability density of the system being in state $x$ at time $t$ given that it was 
in state $x_0$ at time $t_0$:
\[ K(x,t|x_0,t_0) = \int dX P\left[X|x_0,t_0\right] \]
and $K(x,t_0|x_0,t_0) = \delta(x-x_0)$. 
(The notation $\int dX$ here denotes integration over all intermediate states of the trajectory, with the initial and final points fixed.)
Note that $\int dx K(x,t_0|x_0,t_0)$ should be equal to one.
As we are working with a Markov process, one can use the Chapman-Kolmogorov
relations to write a differential equation for $K$:
\[
 \frac{\partial}{\partial t}K(x,t|x_0,t_0) = \int dx' R_t(x'\to x)K(x',t|x_0,t_0)
\]
where $R_t$ is an instantaneous transition rate, given by
\[ R_t(x'\to x) = \lim_{\Delta t\to 0}\frac{K(x,t+\Delta t|x',t) - \delta(x-x')}{\Delta t} \]
and as a consequence $\int dx R_t(x'\to x) = 0$.
We schematically write the differential equation as
\begin{equation}\label{diffK} 
 \frac{\partial}{\partial t}K = L_t K
\end{equation}
where the operator $L_t$ is called the generator of the Markov process.

\paragraph*{Feynman-Kac:}
Now let us consider giving each trajectory a time-dependent statistical weight.
Specifically, we define an arbitrary function $\omega(x,t)$,
and the trajectory-dependent quantity
\begin{equation}
\label{eq:Omega}
\Omega(X) := \int_{t_0}^{t}\omega(x_s,s)ds
\end{equation}
With this, we define a quantity analogous to $K$:
\[ G(x,t|x_0,t_0) := \int dX P\left[X|x_0,t_0\right]e^{\Omega(X)} \]
where the difference lies in the exponential weight assigned to each path.
The goal now is to write down a differential equation for $G$ analogous to (\ref{diffK}).
For a short time interval $dt$ we have
\begin{eqnarray*}
 G(x,t+dt|x',t) &=& \int dX P\left[X|x',t\right]e^{\int_{t}^{t+dt}ds\omega(x_s,s)}\\
&=& e^{\omega(x_t,t)dt}K(x,t+dt|x',t) + o(dt)\\
&=& e^{\omega(x_t,t)dt}[\delta(x-x') +R_t(x'\to x)dt] +o(dt)\\
\end{eqnarray*}
so that
\begin{eqnarray*}
 G(x,t+dt|x_0,t_0) &=& \int dx' G(x,t+dt|x',t)G(x',t|x_0,t_0)\\
&=& (1+\omega(x,t)dt + dtL_t)G(x,t|x_0,t_0)+o(dt)
\end{eqnarray*}
This then finally leads us to:
\begin{equation}\label{diffG}
 \frac{\partial G}{\partial t} = (\omega+L_t)G
\end{equation}
This result is the Feynman-Kac theorem in the present context:
when exponential weights given by equation (\ref{eq:Omega}) are assigned to the trajectories of a process with transition probabilities $K$ satisfying (\ref{diffK}), then the weighted transition probabilities $G$ satisfy (\ref{diffG}).

\paragraph*{Proof of equation (\ref{res3}):}
First note that the probability that the process
is in state $x$ at time $t$ can be written as
\[ f(x,t) = \left< \delta(x-x_t) \right> \]
As this is in our physical systems equal to $\int dx_0 f^{eq}_A(x_0)K(x,t|x_0,t_0)$ for any $t_0<t$, we see that $f$ itself satisfies
\[ \frac{\partial f}{\partial t} = L_t f \]
Recalling that the time-dependence comes from the external work parameter,
we write $L_t = L_{\lambda_t}$. 
We further make the assumption of detailed balance, as above: 
\[ L_{\lambda}e^{-\beta H(x,\lambda)}=0 \]
for every $\lambda$.
The work done during a particular process, up to time $t$, is equal to $w_t(X) = \int_0^t ds \dot{\lambda}\,{\partial H}/{\partial \lambda}$.
Here this work will enter into an exponential weight:
\begin{eqnarray*}
 g(x,t) &:=& \left< \delta(x-x_t)e^{-\beta w_t} \right>\\
&=& \int dx_0 f^{eq}_A(x_0)\int dX P[X|x_0,0]e^{-\beta w_t(X)}\\
&=& \int dx_0 f^{eq}_A(x_0)G(x,t|x_0,0)
\end{eqnarray*}
By the Feynman-Kac theorem we know the differential equation for $G$, and therefore also the one for $g$:
\[ \frac{\partial g}{\partial t} = \Bigl(-\beta\dot{\lambda}\frac{\partial H}{\partial \lambda}+L_t \Bigr)g \]
It can be easily checked that $g(x,t) \propto e^{-\beta H(x,\lambda_t)} $ solves this differential equation,
and because the initial conditions are $g(x,0) = f^{eq}_A(x)$, we finally arrive at the desired result:
\[ g(x,t) = \left< \delta(x-x_t)e^{-\beta w_t} \right> = \frac{1}{Z_A}e^{-\beta H(x,\lambda_t)} \]

\section{Guessing the direction of the arrow of time}
We conclude with a brief discussion of the nature of irreversibility at the microscopic scale,
framed in the context of a thought experiment that involves guessing the chronology of a sequence of events.

Suppose you are shown a movie of a tennis ball falling freely in the absence of significant frictional forces, for instance near the surface of the moon.
If asked whether the movie is shown as it actually happened, or whether it is being run backward in time, you will not be able to answer; the dynamics of the ball is symmetric in time. 
On the other hand, when you are shown a movie of a glass of wine falling
and shattering on the ground, or of the reverse -- shards of broken glass spontaneously assembling themselves and then collecting a stream of wine -- there is no difficulty determining which version depicts the actual sequence of events.
Between these two limiting situations, there are cases like the RNA pulling 
experiment, where frictional forces are certainly non-negligible, but thermal fluctuations complicate the picture.
Let us analyze this situation using the results of the previous section.

As with the examples of the tennis ball and glass of wine, suppose you are shown a microscopic ``movie'' depicting an RNA molecule being stretched, from which you are able to reconstruct the trajectory $X$, as defined earlier.
Note that the corresponding time-reversed trajectory $X^{\dagger}$ depicts a realization of the time-reversed process, involving the contraction of the RNA molecule.
From the data -- that is, the movie -- you are asked to guess the direction of time's arrow:
does the movie show the sequence of events as they actually occurred (e.g. stretching) or are you instead observing a movie that was filmed during the reverse process (contraction), now run backward in time?
Let $F$, for ``forward'', and $R$, for ``reverse'', denote these two possibilities, respectively.

\begin{figure}[ht]
 \includegraphics[width = 10cm]{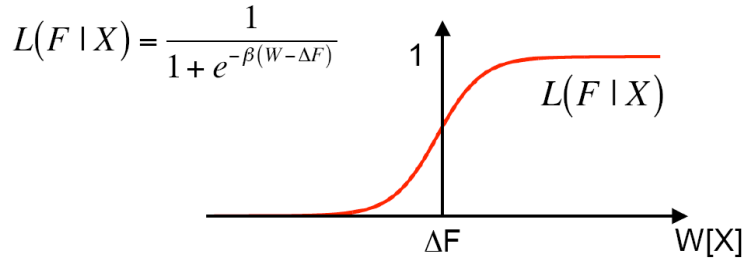}
\caption{A graphical depiction of $L(F|X)$.}
\label{timesarrow}
\end{figure}

Since both $X$ and $X^\dagger$ represent perfectly valid microscopic trajectories, there is no definitive answer to the above question.
Rather, the problem becomes an exercise in likelihood estimation: given the data ($X$, equivalently $X^\dagger$) which of the two hypotheses, $F$ or $R$, is more likely?
The answer is provided by Bayes' rule in probability theory:
assuming we have no {\it a priori} reason to favor one hypothesis over the other, the likelihood $L(F|X)$ of the forward hypothesis, given the observed data $X$, is proportional to the probability to generate that data when performing the forward process:
$L(F|X) \propto P^F(X)$.
The analogous statement holds in the reverse case: $L(R|X^\dagger) \propto P^R(X^\dagger)$.
Combining these with a statement of normalization, namely $L(F|X) + L(R|X^\dagger) = 1$, we conclude that
\begin{eqnarray*}
 L(F|X) = \frac{P^F(X)}{P^F(X)+P^R(X^{\dagger})}
\end{eqnarray*}
We now invoke equation (\ref{tussenres}) to arrive at the result,
\begin{equation}
 L(F|X) = \frac{1}{1+ e^{-\beta(W-\Delta F)}}
\end{equation}
This result has previously been derived, with a somewhat different intepretation, in the context of free energy estimation \cite{mar07,shi03}.
Recall that $W$ is the work done during the process $X$. In Figure \ref{timesarrow} we see
a graphical depiction of $L(F|X)$.
When $W>\Delta F$, it is more likely that we saw the movie in the right order -- that is, the molecule was actually stretched --
while for $W<\Delta F$, it is more likely that we saw the movie in reverse. 
The transition from $L \sim 0$ to $L \sim 1$ occurs over an interval of $W$ of width $\sim k_BT$.
Thus unless the work is within a few $k_BT$ of the free energy difference, we can determine the direction of the arrow of time with near certainty.
This is entirely consistent with our observation that we have no problem distinguishing the direction of time's arrow for macroscopically irreversible processes like that shattering of a glass of wine.

\section*{Acknowledgments}

We are grateful to the organizers of FPSPXII for arranging a highly enjoyable and successful summer school.
C.J. acknowledges financial support from the National Science Foundation (USA), under CHE-0841557.
B.W acknowledges support as Aspirant of FWO, Flanders.

\appendix

\section{Jensen's inequality}

Take a space $\Omega$ of configurations, and an arbitrary probability density $p(x)$ on that space.
Denote averages of functions $f$ with respect to $p$ by $\left<f\right> = \int dx p(x)f(x)$.
Now consider a convex function $f(x)$.
By the definition of convexity, this function satisfies
\[ f(x) \geq f(a)+ (x-a)f'(a) \]
for any $x,a\in\Omega$.
Equivalently, we can write
\[ f(g(x)) \geq f(a) + (g(x)-a)f'(a) \]
for an arbitrary (not necessarily convex) function $g(x)$.
Multiplying both sides by $p(x)$ and integrating over $\Omega$ we get
\begin{equation}
 \left<f(g)\right> \geq f(a) + (\left<g\right> -a)f'(a)
\end{equation}
If we now choose $a$ such that $a = \left<g\right>$, we arrive at Jensen's inequality:
\[ \left<f(g)\right> \geq f(\left<g\right>) \]
For the special case $g(x) = x$, this becomes
\[ \left<f(x)\right> \geq f(\left<x\right>) \]


\end{document}